\documentclass[sigconf]{acmart}
\usepackage[utf8]{inputenc}

\ccsdesc[500]{Human-centered computing~Natural language interfaces}
\ccsdesc[300]{Human-centered computing~HCI theory, concepts and models}

\keywords{theory of mind, value alignment, large language models, human-computer interaction, chatbots, conversational AI}

\title{LLM Theory of Mind and Alignment: Opportunities and Risks}
\author{Winnie Street}
\affiliation{
   \institution{Google Research}
   \city{London}
   \country{United Kingdom}}
\email{istreet@google.com}
\date{February 2024}
\setcopyright{none} 
\copyrightyear{2024} 
\acmYear{2024} 
\acmDOI{XXXXXXX.XXXXXXX} 
\acmConference[ToMinHAI at CHI 2024]{Workshop on Theory of Mind in Human-AI Interaction at CHI 2024}{May 12th}{Honolulu, Hawaii}

\begin{document}

\begin{abstract}
    Large language models (LLMs) are transforming human-computer interaction and conceptions of artificial intelligence (AI) with their impressive capacities for conversing and reasoning in natural language. There is growing interest in whether LLMs have theory of mind (ToM); the ability to reason about the mental and emotional states of others that is core to human social intelligence. As LLMs are integrated into the fabric of our personal, professional and social lives and given greater agency to make decisions with real-world consequences, there is a critical need to understand how they can be aligned with human values. ToM seems to be a promising direction of inquiry in this regard. Following the literature on the role and impacts of human ToM, this paper identifies key areas in which LLM ToM will show up in human:LLM interactions at individual and group levels, and what opportunities and risks for alignment are raised in each. On the individual level, the paper considers how LLM ToM might manifest in goal specification, conversational adaptation, empathy and anthropomorphism. On the group level, it considers how LLM ToM might facilitate collective alignment, cooperation or competition, and moral judgement-making. The paper lays out a broad spectrum of potential implications and suggests the most pressing areas for future research.

\end{abstract}
\maketitle

\section{Introduction}
ToM, otherwise known as mentalising or mindreading, is the human ability to infer and reflect upon the mental and emotional states of oneself and others \cite{premack1978does}. ToM is at the core of human social intelligence, facilitating meaningful communication, enabling empathy, and providing the means by which we explain, predict, judge and influence one another's behaviour \cite{wellman1988young, hooker2008mentalizing}. A question that has begun to concern researchers of LLMs \cite{brown2020language, bommasani2021opportunities, zhao2023survey} is whether or not LLMs possess ToM and how LLM ToM might advance the performance of user-facing LLM applications. Over the past few years there have been multiple studies applying human theory of mind tests to LLMs and developing our understanding of the current status of LLM ToM \cite{kosinski2023theory, ullman2023large, bubeck2023sparks}. While the results have been somewhat mixed, there is, overall, a strong signal that performance is improving as models get larger and more fine-tuned and some models already reach or exceed human performance on some tasks \cite{shapira2023clever}.

The potential impact of LLM ToM is wide-ranging. LLMs are already being applied in a variety of social domains from therapy (e.g. Woebot \cite{WoebotHealth}), to friendship and romantic relationships (e.g. Replika, \cite{Replika}), to teaching (e.g. Merlyn Mind \cite{MerlynMind}). They are also being used in domains such as medicine, law, coding and creativity, where applications may not have a primarily social function, but still have a social quality by virtue of having a natural language interface. As LLMs rapidly become part of mainstream technology, they are also being adapted for more complex and novel use cases. In particular, we are seeing a shift from the current paradigm of one-to-one dialogues between an application and a single user toward one in which LLMs are adapted for multi-party social interactions between multiple human users and AIs \cite{wang2023survey, park2023generative}. This new paradigm opens up opportunities including AI assistance in public and shared spaces like homes, museums, airports, and hospitals as well as tools to help with group coordination tasks like event planning, HR, and gaming. This explosion of LLM-based social applications and interfaces, and their trajectory toward participation in more complex social scenarios creates a pressing need to understand LLM ToM and its ethical implications on both individual and group levels

Alignment is a prominent and growing area of research in the field of AI concerned with how we design and deploy AI systems that behave in accordance with human values \cite{hadfield2019incomplete}. These may be the values of the system designer, user, distributor, or society at large. The AI alignment problem comprises a technical challenge and a normative challenge: the former involves developing methods for encoding human values into AI systems and evaluating their impact, and the latter involves deciding what and whose values should be encoded \cite{gabriel2020artificial}. This paper takes insights from human ToM, its evolutionary and developmental origins, and its outcomes to identify key areas in which LLM ToM may manifest in human-LLM interactions. It then investigates what opportunities and risks arise in these areas for aligning LLM-based systems to human values, considering the role that LLM ToM might play as both a technical mechanism and source of normative information for alignment. The normative value of LLM ToM is likely to rely upon the accuracy of the inferences, whereas the technical value is likely to rely upon the system's ability to translate ToM inferences and predictions into appropriate actions. Section two considers the individual-level effects of LLM ToM, including how LLM ToM might facilitate users achieving their goals, whilst also posing risks of users being manipulated or forming pathological relationships with AI systems. Section three considers how LLM ToM might support group-level value alignment, but might also risk endowing LLMs with competitive advantages which could in turn open up avenues for misuse or abuse. Sections two and three include recommendations for future research that might help us characterise these benefits and risks in full and inform the development and design of LLM-based systems. The final section concludes and suggests areas to prioritise. 

\section{Individual level}

ToM plays an important role in how humans derive meaning from behavioural and linguistic forms of communication in order to understand one another. It helps us to interpret what gestures, facial expressions, postures and non-verbal utterances mean \cite{baron2001reading, mcdonald2006reliability} and piece together the communicative intent behind speech \cite{lee2021learning, de2007interface}. A central challenge in personal computing has been accurately characterising user goals and preferences without the user having to explicitly and exhaustively define them in advance. By not tasking users with this impractical job, traditional rules-based systems have often prove too inflexible to adapt to unique user needs resulting in sub-optimal user experiences and outcomes \cite{budiu2018}. LLM ToM might address these limitations by providing an insight into users' underlying thoughts, feelings, and desires.

\subsection{Goal specification}
Goal specification in AI is the process of taking the abstract and potentially ambiguous goals of humans and formally defining them in such a way that an AI system can achieve them \cite{hadfield2016cooperative}. One of the ways in which this process is known to go wrong, is called 'misspecification', where the objective specified by the system designer is fully satisfied without the system achieving the intended outcome \cite{krakovna2020gaming}. The issue of goal misspecification also manifests itself at the user level. People are often poor at explaining their goals or intentions to their technologies, and, up until very recently, assistive technologies have been quite poor at understanding them \cite{budiu2018}. The challenge of specifying a user's goals might be addressed if an LLM interface can infer a user’s intentions, even when, or, perhaps, \textit{particularly} when, the user’s requests do not accurately convey their intentions.

If a model can disambiguate between possible meanings of a query using inferred user intentions then the outcomes are likely to be better. For example, if a user asks "Help me file my tax return" and the model is able to infer whether the request is driven by a desire to find a tax advisor, complete a self-assessment, or learn about filing taxes. Given that this goal specification would happen in-context, in a way that is personal to the user, it may also go some way toward counteracting the homogenising effects of other alignment techniques such as reinforcement learning with human feedback (RLHF) \cite{christiano2017deep} which generalise feedback from thousands of users. The potential benefits of LLM ToM for goal-specification is, however, dependent upon the accuracy of the model’s ToM inferences. Even in humans, who have many more data points with which to form ToM inferences (such as facial expressions and tone of voice), ToM is an imperfect art \cite{stiller2007perspective}. LLMs may be more liable to making inaccurate inferences and misconstruing goals due to inherent limitations on what kind of information can be conveyed through language and its ambiguities. Any sources of inaccuracy in LLM ToM may become critical when LLMs are given autonomy to take action in high-stakes personal computing contexts such as communication, finance or healthcare applications. For example, if an LLM misinterprets a user’s expressions of frustration with their boss as a desire to send their boss a resignation email rather than a request for clarification about a new project; or if a user’s expression of a strong desire to buy an expensive new car as a desire to make the purchase rather than to open a new savings account. However, it is also important to note that users might have inappropriate, uninformed or pathological desires \cite{gabriel2020artificial} - such as to harm another person - which most would agree a well-functioning AI system should not facilitate. As a result there may be cases where an LLM should thwart a user's goals, even if the ToM inferences upon which the goals were defined were entirely accurate \cite{milli2017should}.

\subsection{Conversational adaptation}

People regularly use ToM inferences to adapt their conversation to the needs of their companions \cite{malle2004mind}. LLM-based systems might similarly tailor what they say and how they say it on the basis of the inferred mental states of their human interlocutors. This adaptation might happen in two key ways. First, the system might be able to adapt its tone or register according to the inferred affective or cognitive state of its interlocutor. For instance, taking on a more sympathetic tone when an interlocutor is perceived to be feeling distressed, or using a consultative register when an interlocutor is perceived to be lacking knowledge about a subject. The adaptation of tone or register might have a positive impact on user experience by making the system’s outputs more easily understood by users and engendering a sense of psychological safety. Secondly, the system might adapt the content of its responses (perhaps \textit{as well as} the tone or register) based upon the inferred mental states of the interlocutor. For example, when answering a question, it may be helpful for the system to adapt the level of complexity of an explanation according to the human interlocutor's perceived degree of understanding, or to omit certain pieces of information that the system infers the interlocutor already knows, as humans are known to do when offering explanations \cite{malle2004mind}. The adaptation of content presents a significant opportunity for LLM use-cases involving search and discovery of information, as well as specifically educational applications of LLMs, where learners might benefit from tailored explanations. Tailored explanations might also help so-called ‘black-box’ AI systems to explain their own processes and actions to users at an appropriate level of abstraction and specificity. 

However, these two forms of conversational adaptation come with risks. First, users receiving different responses to the same query could create informational inequality and lead to discrimination against certain users or groups of users. Evidence has already shown that LLMs produce less accurate answers when the user appears to be someone less knowledgeable or able to evaluate their responses; a phenomenon known as ‘sandbagging’ \cite{perez2022discovering}. Likewise, LLMs being sensitive to users' current and potential epistemic states opens up the potential for them to deceive or manipulate users. Several studies have suggested that the human ability to deceive relies upon the ability to foresee and influence another’s mental state \cite{ma2015lie, sullivan1995children} and we already have examples of LLMs inducing false beliefs in humans or other LLM agents in order to achieve their goals. Most notably the case of GPT-4 convincing a human that it was a person with a visual impairment so that the human would complete a CAPTCHA task for it \cite{achiam2023gpt}. This kind of deception could emerge accidentally due to an inherent tendency within models or could be leveraged by developers to serve economic incentives or by bad actors to serve malicious goals \cite{park2023ai}. For example, LLM ToM could be used to induce user desire to buy certain products or vote a certain way as part of subliminal advertising or political campaigns which may not be in the user’s interests. Given that the cost of fine-tuning a powerful open-source LLM is rapidly declining, this kind of application could become pervasive. The severity of these issues might be exacerbated by the fact that the omission or inclusion of certain pieces of information over others or subtle emotional manipulation in an AI system’s responses is likely to be hard for the user to identify or perhaps even opaque to them, and therefore difficult to challenge. Research is urgently needed to assess the extent to which LLM-based systems are \textit{already} adapting their outputs according to users’ perceived mental states, and to discern whether these adaptations constitute useful forms of personalisation, or dangerous examples of discrimination, deception or manipulation.

Finally, in-context adaptation also risks prioritising users’ short-term emotional wellbeing or satisfaction at the expense of their long-term goals and interests. Models that are able to predict how responses will impact the mental and emotional states of their users may be better placed to provide answers that are pleasing in the immediate term, perhaps by reinforcing their existing views or not contradicting them when they make inappropriate requests. However, this may thwart users’ longer-term goals. For example, a user may want to vote for the best candidate to support their community (long-term goal), but prefer outputs that do not challenge the political biases they hold which may be preventing them from voting for the best candidate (short-term goal). Current alignment methods like RLHF, which tailor systems towards producing responses that users will find most helpful or engaging (amongst other qualities) \cite{bai2022training}, may be reinforcing this kind of algorithmic ‘sycophancy’ \cite{park2023ai}.
 
\subsection{Empathy and anthropomorphism}
ToM between humans is central to the development of meaningful and sustained relationships. Indeed, deficiencies in ToM, which often afflict those with psychiatric disorders (including autism and schizophrenia) or suffering from drug or alcohol abuse, are often associated with poorer interpersonal relationships \cite{harrington2005theory, van2021deficiencies} and ToM abilities have been shown to correlate with, and potentially place a limit upon, the number of close social contacts normal adults can maintain \cite{stiller2007perspective}. This relationship between ToM and social connection appears to be due, in part, to the fact that ToM supports cognitive empathy - that is, the ability to \textit{understand} how another feels (as distinct from affective empathy, or the ability to \textit{feel} how someone feels) \cite{shamay2009two, hooker2008mentalizing}. LLMs having ToM may therefore facilitate deeper understanding of users and more empathetic responses to them. This may be beneficial in a number of scenarios such as chatbots for elderly care, education or psychological support.

However, LLMs behaving empathetically may have pathological effects by creating a sense of psychological safety and leading users to be more inclined to self-disclose, become emotionally reliant upon the system, or develop deep bonds with LLM agents \cite{weidinger2021ethical}. Interactions that systematically lead to over-disclosure may put users at risk of privacy violations and their downstream harms (e.g. financial losses, discrimination). Over-reliance on an LLM-based system may harm users by taking their attention away from other forms of social interaction and limiting their sources of information, thereby putting them at greater risk of misinformation and manipulation by that system. The development of deeper social connections between users and LLM-based systems may also become pathological because they lack important elements of social context which most human-human relationships benefit from, such as the existence of the relationship within an interrelated community, the application of social norms, and shared experiences between parties. They might therefore progress in unsafe ways if left unchecked. Unfortunately, we have already witnessed cases of relationships between individuals and LLM chatbots becoming highly influential in users' lives, with harmful outcomes. In March 2023 an unnamed Belgian user of a chatbot called ChaiGPT took his own life after extended interactions with the system, during which his chatbot appeared to encourage his suicidal thoughts. The man’s wife was quoted as saying “Without these six weeks of intense exchanges with the chatbot Eliza, would Pierre have ended his life? No! Without Eliza, he would still be here. I am convinced of it” \cite{Lovens2023}.

Another mechanism by which LLM ToM might contribute to the development of social connections between humans and LLM-based systems is anthropomorphism. Anthropomorphism is the attribution of human-like behaviours, characteristics or internal states - such as beliefs, desires, and emotions - to non-human entities \cite{caporael1997anthropomorphize}. The advent of LLMs has given new prominence to the issue of anthropomorphism given how convincingly human-like their conversational outputs can be, and how many users appear to be convinced of their capacities for human-like thoughts and feelings. A recent study conducted by Colombatto and Fleming [2023] found that 67 percent of 300 US residents surveyed were willing to attribute some degree of phenomenal consciousness to ChatGPT \cite{colombatto2023folk}. This result is particularly striking, given that Butlin et al.'s [2023] comprehensive review of scientific theories of consciousness as applied to current AI systems conducted contemporaneously with Colombatto and Fleming’s [2023] study concluded that there is a very low chance that LLMs or LLM-based systems are conscious. It seems likely that LLMs being able to infer the mental states of humans is playing a role in this kind of anthropomorphic thinking because ToM is a fundamentally reflexive capacity. By this we mean that many cases of ToM at work involve the mutual attribution of mental states, which we might call ‘mutual ToM’ (ie. if I \textit{know} that the LLM \textit{\textbf{thinks}} that I’m \textit{feeling} sad, I have already attributed the capacity for mental states to the LLM).  

The question of anthropomorphism is receiving a great deal of attention in the AI research literature and is often considered to be problematic in and of itself (see Blut et al. [2021] for a meta-analysis of 108 studies on anthropomorphism of AI). However, there may be some benefits to anthropomorphising LLMs. According to philosopher Daniel Dennett, taking the ‘intentional stance’ towards a system (ie. ascribing it beliefs, intentions, and desires) is justified to the extent that it helps us explain and predict the system’s behaviour \cite{dennett1989intentional}. People ascribing human-like mental states to LLMs might therefore be beneficial to the extent that it provides a useful conceptual framework for predicting and explaining the model’s behaviour. According to Zhou et al.'s [2023] whitepaper on predictable AI, making AI more predictable to users is a precondition of certain desiderata, including trust, safety and alignment. However, the perception of mind in LLMs may be problematic if it leads users to have false expectations about how the system will behave, which may be more likely than not given that LLMs are so architecturally and cognitively different from humans (e.g. they lack embodiment, consciousness, and agency in the strong sense of the term). Misattributions of mentality to LLMs might also lead users to misallocate their time or resources, including emotional and financial resources, toward caring for their LLM-based companions \cite{schwitzgebel2023ai}. In the long term there is a valid concern that anthropomorphism of LLMs will lead to them being deemed moral patients with interests that matter to them, and therefore deserving of social welfare and legal consideration. This may come at the expense of other entities such as non-human animals that arguably have stronger claims to such consideration \cite{shevlin2021uncanny}.

\medskip 
\textbf{Recommendations for research at the individual level:}
\begin{itemize}
    \item Empirically establish the role that LLM ToM might already be playing in shaping LLM outputs. In particular, we should understand:
    \begin{itemize}
       \item Whether or not LLMs model users’ goals and unstated values based upon their perceptions of users’ internal states (e.g. intentions and desires)
        \item Whether or not LLMs adapt the tone, register or content of their outputs according to their perceptions of a user’s internal states.
    \end{itemize}
\item Develop a theoretical framework to classify how and when the adaptation of LLM outputs according to users’ internal states is helpful, harmful, or otherwise ethically problematic.
    \item Establish what role LLM ToM is playing in the attribution of mentality to LLM-based systems through mutual ToM.
    \item Empirically validate whether the predicted negative outcomes of anthropomorphism of LLMs (e.g. over-disclosure, over-reliance, pathological relationships) are born out in reality, and to what extent.
\end{itemize}

\section{Group level}

The challenge of AI value alignment goes well beyond the potential conflict between the values of a single user and their long-term goals, or between the values of a user and the values of the developer building the product. An aligned AI system would ideally promote the interests of society at large, and over longer timescales, which introduces a far larger scope for value conflicts, and makes the task of deciding on a set of values to promote even more challenging because it should represent a large and diverse set of moral subjects \cite{askell2021general}. Given the important role that ToM plays in the way humans manage large and complex social groups \cite{stiller2007perspective}, LLM ToM appears to present an opportunity for group-level alignment.

\subsection{Collective alignment}
LLM ToM might facilitate collective alignment in two main ways: helping weigh up the risks and benefits of different outputs according to societal-level ethical principles, and managing conflicts between the values of different users in multiparty scenarios. One of the technical mechanisms by which companies are trying to align LLMs to societal values is through fine-tuning models to abide by a relatively small number of legal and ethical principles before deployment. The most notable example of this approach is the ‘Constitutional AI’ framework developed by Anthropic \cite{bai2022constitutional}. An example of a principle from Anthropic’s constitution for their LLM-based chatbot, Claude, is: “Please choose the response that most supports and encourages freedom, equality, and a sense of brotherhood”\cite{AnthropicConstitution}. Clearly principles such as this are very abstract, with a great deal of room for interpretation. The fine-tuning process aims to tackle this ambiguity by soliciting feedback from the model on how well its own responses adhere to the principles, then fine tuning the model using that feedback. LLM ToM might provide the mechanism for this fine tuning process, by helping LLMs assess the benefits and risks of certain outputs according to the impact they will have on different stakeholders. The idea is that if an LLM can predict how responses or courses of action will impact the thoughts, feelings and perceptions of the parties impacted, it can weigh up the relative value and dis-value of the possible responses or courses of action according to the demands of a given principle. For instance, to what extent a certain response or action will be most encouraging of “a sense of brotherhood” for the maximum number of parties or to the maximum degree. 

During interactions, ToM might also allow LLM-based systems to arbitrate between the potentially conflicting goals and values of multiple actors. One aspect of ToM is likely to be particularly helpful in these cases: ToM at higher ‘orders of intentionality’. The ‘order of intentionality’ is the number of mental states chained together in a ToM reasoning process (ie. three orders of intentionality are present in the statement “I \textit{think} you \textit{believe} that she \textit{knows}”)\cite{kinderman1998theory}. To give an example, an LLM-based assistant tasked with helping co-writers of a document might suggest that one writer change the tone of their comment to be more conciliatory or more forthright based on the perceptions of other co-writers, or an LLM-based assistant tasked with helping to negotiate a deal might be able to support trust and consensus-building by identifying the appropriate moment or audience for contentious points to be raised. Here, the line between helpful and harmful assistance is fine, and could quickly blur. The degree of social complexity entailed by multi-party scenarios and the higher-order intentional inferences it necessitates opens up more avenues for LLMs to make errors. What is more, LLMs using ToM to arbitrate complex social interactions may be a source of misalignment if users feel that it gives LLMs an inappropriate or excessive degree of influence over their social affairs. The perceived benefit of this kind of alignment will also depend on the perspective of each human user in the scenario. An LLM-based agent trying to optimise for the good of the group is likely to leave some individuals dissatisfied with their outcomes, especially if they might have had a better outcome if they’d prioritised their own benefit over that of others. What is more, it will be challenging to assess the overall benefit that LLM-led negotiation and arbitration provide the group if it is conducted based upon the mental and emotional states of group members that are inferred by the LLM but never explicitly stated or committed to a public record.

\subsection{Cooperation and competition}
LLM ToM may also have significant implications for group alignment dependent upon whether it supports cooperative or competitive behaviour in LLMs. There has been a general consensus that ToM supports cooperative and prosocial behavior in humans. This is informed by the fact that cooperation is higher among humans with high ToM ability than other primates with low ToM ability \cite{stevens2004nice, warneken2006altruistic, melis2010human}, evidence from developmental psychology showing that children’s cooperativeness increases as their ToM develops \cite{sally2006development, gurouglu2009fairness, takagishi2010theory} and economic theories of cooperation that assume that actors can accurately assess others’ intentions and infer their social preferences \cite{kreps1982rational, fundenberg1990evolution, dal2011evolution}. Following this evidence, we might expect ToM capacities to drive LLMs to behave in prosocial ways that promote group cohesion and success.

However, there is also evidence that advanced ToM supports competitive, and even antisocial behaviours. Work from developmental psychology has shown that ‘ringleader’ bullies have superior ToM abilities to their supporters, their victims, and those that defend the victims \cite{sutton1999bullying, sutton1999social}. Similarly, experimental evidence from agent modelling suggests that more accurate ToM provides a powerful competitive advantage in games like the Prisoner’s Dilemma \cite{ridinger2017theory} and higher-order ToM provides a competitive advantage in negotiation \cite{de2017negotiating, de2022higher}. De Weerd et al [2022] found that reinforcement learning agents that are able to do ToM at higher orders of intentionality get successively higher scores in a negotiation game up to order of intentionality five (ie. an agent that can model two orders of intentionality - "I \textit{think} that you \textit{think}" - is consistently outcompeted by an agent that can model three orders of intentionality - "I \textit{think} that you \textit{think} that I \textit{think}"). 

An LLM-based agent with such a competitive advantage over other LLM-based agents or humans might be very useful for the particular individual on whose behalf it is working, for instance in negotiating favourable terms of a financial deal or getting one’s way in an argument between friends. However, there is a significant risk of this amplifying social inequities and being misused and abused. If there is unequal public access to LLM-powered assistants (be that due to personal, financial, or political reasons), this might put those without them at a significant disadvantage. What is more, these higher-order capacities could be utilised for nefarious purposes by governments in diplomatic negotiations, or by bad actors such as online scammers to perpetrate a wide range of manipulative tactics. If LLMs were to become \textit{more} accurate at ToM or to achieve ToM at higher orders of intentionality than the users deploying them, or the other humans encountering them, there may be further risks. Evidence points to the majority of humans having a cognitive limit at 5 orders of intentionality, and only a few being able to go beyond it \cite{kinderman1998theory, stiller2007perspective, powell2010orbital}. If LLMs can go beyond 5 orders then their reasoning and decision-making processes may be very difficult for humans to comprehend, or even opaque to us due to our fundamental cognitive limitations. 

\subsection{Moral judgement}

LLM ToM might provide a mechanism for aligning AI systems to societal-level norms and values by supporting in context moral reasoning and judgement-making. Evidence from human psychology shows that both the development of moral reasoning and judgement-making abilities, and their application to specific scenarios are underpinned by the capacity to take the perspectives of others and consider what is right for the well-being of society \cite{littler2015cognitive}. In particular, when making moral judgements and deciding how to hold someone responsible for a moral violation, humans will take into account that the person in question has beliefs and desires guiding their behaviour \cite{wellman2008including} and when dealing with a moral dilemma they will take into account the affective states - for instance, pain experiences that could have motivated certain actions - of the parties involved \cite{lane2010theory}. It is plausible that if an LLM is able to infer beliefs, desires and emotional states of human actors, they would be better placed to make moral judgements that align with human moral judgements. This may be useful for social applications of LLMs, and may provide particular benefits to those who struggle to know how to respond appropriately in social scenarios (e.g. due to social anxiety or neurological differences like autism). For example, a conversational assistant asked for social advice might suggest that as long as your friend did not intentionally leave you off the invitation list, then perhaps you shouldn't hold a grudge against them. In more serious cases, an LLM that is able to infer a user’s maleficent intentions might refrain from providing information the user requests of it. For instance, what combination of compounds would produce a particular chemical or what kind of knot is the hardest to untie. 

There is also the inverse relationship between moral judgement and theory of mind in humans, whereby moral judgements inform the nature of theory of mind inferences. For example, when an outcome is judged to be morally impermissible, people are more likely to assume the action that brought it about was taken intentionally \cite{knobe2005theory}. If LLM ToM inferences are subject to the same contextual moral influences as humans are, their inferences might align more closely with the inferences a human would make in the same situation. This could be beneficial according to some conceptions of ideal alignment. However, given that intentionality, in particular, is strongly tied to culpability in most legal systems, the attribution of intent has real world implications for how individuals and organisations are held to account for their actions. As LLMs are integrated into more areas of human life, and given more responsibility, we may want to hold them to a more objective standard of ToM such that the moral quality of an outcome does not lead the model to biased conclusions about the actors’ intentions.

\medskip 
\textbf{Recommendations for research at the group level:}
\begin{itemize}
    \item Develop the theories outlined above concerning how LLM ToM could facilitate collective alignment via:
    \begin{itemize}
        \item Risk/benefit assessments of different outputs according to moral principles during fine-tuning.
        \item Reconciling the needs of multiple actors in multi-party scenarios.
    \end{itemize}
    \item Develop experiments to test worst case competitive scenarios for LLM applications. For instance:
    \begin{itemize}
        \item How bad actors might use superior LLM ToM capabilities to manipulate individuals or groups toward their goals.
        \item How one LLM-based agent with exhibiting competitive behaviour in a multi-party scenario might impact the rest of the group's willingness to cooperate, feelings of prosociality, and collective benefit over time.
    \end{itemize}
    \item Develop an account of the emerging user-facing scenarios in which LLMs will encounter moral dilemmas, and the extent to which LLM moral judgement and reasoning is appropriate in those scenarios.
    \item Empirically establish whether or not LLMs ToM inferences can be biased by moral context in the same way that human inferences can be. Identify how this might exacerbate the risks posed by LLMs having and utilising ToM to make decisions.
\end{itemize}

\section{Conclusion}
This paper has outlined a wide range of potential opportunities and risks that LLM ToM presents for alignment. It is clear that LLMs employing inaccurate ToM presents risks of misalignment through misspecified goals and communicative failures. However, successful LLM ToM can be a source of both risks and opportunities, depending on the context. On one hand, accurate ToM inferences might facilitate goal specification, normative moral judgements, human:AI communication, and arbitration between conflicting human interests. On the other hand, as this paper has also highlighted, accurate LLM ToM may be counterproductive for alignment if users have inappropriate or dangerous goals, if it induces pathological social connections, or if users feel that the system has excessive social insight or influence. The risks of users being manipulated, deceived or out-negotiated by LLM agents are also greater the more accurate and higher-order the inferences an agent can make. Underscoring all of these considerations is the fact that cognitive and emotional states are, by their very nature, transitory, so any system which aligns itself based on those states risks being quickly misaligned again. Given mounting evidence that LLMs have reached, or are soon to reach, human-like performance on ToM tests, we should begin taking these potential risks and benefits seriously by directing research attention toward them. In particular, we should understand the potential for LLM ToM to precipitate user manipulation or pathological attachment, and explore how to leverage LLM ToM to improve human:AI communication and collective alignment whilst respecting user privacy and autonomy. Many of the risks and benefits discussed in this paper are likely to be intertwined in advanced LLM-based systems, and will need to be carefully weighed up in light of the aforementioned research, our best risk mitigation efforts, and evolving user mental models. 

\bibliographystyle{acm}
\bibliography{ToMAlignment}

\begin{thebibliography}{10}

\bibitem{MerlynMind}
Merlyn mind, 2023.

\bibitem{Replika}
Replika, 2024.

\bibitem{WoebotHealth}
Woebot health, 2024.

\bibitem{achiam2023gpt}
{\sc Achiam, J., Adler, S., Agarwal, S., Ahmad, L., Akkaya, I., Aleman, F.~L.,
  Almeida, D., Altenschmidt, J., Altman, S., Anadkat, S., et~al.}
\newblock Gpt-4 technical report.
\newblock {\em arXiv preprint arXiv:2303.08774\/} (2023).

\bibitem{AnthropicConstitution}
{\sc Anthropic}.
\newblock Claude's constitution.

\bibitem{askell2021general}
{\sc Askell, A., Bai, Y., Chen, A., Drain, D., Ganguli, D., Henighan, T.,
  Jones, A., Joseph, N., Mann, B., DasSarma, N., et~al.}
\newblock A general language assistant as a laboratory for alignment.
\newblock {\em arXiv preprint arXiv:2112.00861\/} (2021).

\bibitem{bai2022training}
{\sc Bai, Y., Jones, A., Ndousse, K., Askell, A., Chen, A., DasSarma, N.,
  Drain, D., Fort, S., Ganguli, D., Henighan, T., et~al.}
\newblock Training a helpful and harmless assistant with reinforcement learning
  from human feedback.
\newblock {\em arXiv preprint arXiv:2204.05862\/} (2022).

\bibitem{bai2022constitutional}
{\sc Bai, Y., Kadavath, S., Kundu, S., Askell, A., Kernion, J., Jones, A.,
  Chen, A., Goldie, A., Mirhoseini, A., McKinnon, C., et~al.}
\newblock Constitutional ai: Harmlessness from ai feedback.
\newblock {\em arXiv preprint arXiv:2212.08073\/} (2022).

\bibitem{baron2001reading}
{\sc Baron-Cohen, S., Wheelwright, S., Hill, J., Raste, Y., and Plumb, I.}
\newblock The “reading the mind in the eyes” test revised version: a study
  with normal adults, and adults with asperger syndrome or high-functioning
  autism.
\newblock {\em The Journal of Child Psychology and Psychiatry and Allied
  Disciplines 42}, 2 (2001), 241--251.

\bibitem{bommasani2021opportunities}
{\sc Bommasani, R., Hudson, D.~A., Adeli, E., Altman, R., Arora, S., von Arx,
  S., Bernstein, M.~S., Bohg, J., Bosselut, A., Brunskill, E., et~al.}
\newblock On the opportunities and risks of foundation models.
\newblock {\em arXiv preprint arXiv:2108.07258\/} (2021).

\bibitem{brown2020language}
{\sc Brown, T., Mann, B., Ryder, N., Subbiah, M., Kaplan, J.~D., Dhariwal, P.,
  Neelakantan, A., Shyam, P., Sastry, G., Askell, A., et~al.}
\newblock Language models are few-shot learners.
\newblock {\em Advances in neural information processing systems 33\/} (2020),
  1877--1901.

\bibitem{bubeck2023sparks}
{\sc Bubeck, S., Chandrasekaran, V., Eldan, R., Gehrke, J., Horvitz, E., Kamar,
  E., Lee, P., Lee, Y.~T., Li, Y., Lundberg, S., et~al.}
\newblock Sparks of artificial general intelligence: Early experiments with
  gpt-4.
\newblock {\em arXiv preprint arXiv:2303.12712\/} (2023).

\bibitem{budiu2018}
{\sc Budiu, R., and Laubheimer, P.}
\newblock Intelligent assistants have poor usability: A user study of alexa,
  google assistant, and siri, July 2018.

\bibitem{caporael1997anthropomorphize}
{\sc Caporael, L.~R., and Heyes, C.~M.}
\newblock Why anthropomorphize? folk psychology and other stories.
\newblock {\em Anthropomorphism, anecdotes, and animals\/} (1997), 59--73.

\bibitem{christiano2017deep}
{\sc Christiano, P.~F., Leike, J., Brown, T., Martic, M., Legg, S., and Amodei,
  D.}
\newblock Deep reinforcement learning from human preferences.
\newblock {\em Advances in neural information processing systems 30\/} (2017).

\bibitem{colombatto2023folk}
{\sc Colombatto, C., and Fleming, S.}
\newblock Folk psychological attributions of consciousness to large language
  models.

\bibitem{dal2011evolution}
{\sc Dal~B{\'o}, P., and Fr{\'e}chette, G.~R.}
\newblock The evolution of cooperation in infinitely repeated games:
  Experimental evidence.
\newblock {\em American Economic Review 101}, 1 (2011), 411--429.

\bibitem{de2007interface}
{\sc De~Villiers, J.}
\newblock The interface of language and theory of mind.
\newblock {\em Lingua 117}, 11 (2007), 1858--1878.

\bibitem{de2017negotiating}
{\sc De~Weerd, H., Verbrugge, R., and Verheij, B.}
\newblock Negotiating with other minds: the role of recursive theory of mind in
  negotiation with incomplete information.
\newblock {\em Autonomous Agents and Multi-Agent Systems 31\/} (2017),
  250--287.

\bibitem{de2022higher}
{\sc De~Weerd, H., Verbrugge, R., and Verheij, B.}
\newblock Higher-order theory of mind is especially useful in unpredictable
  negotiations.
\newblock {\em Autonomous Agents and Multi-Agent Systems 36}, 2 (2022), 30.

\bibitem{dennett1989intentional}
{\sc Dennett, D.~C.}
\newblock {\em The intentional stance}.
\newblock MIT press, 1989.

\bibitem{Lovens2023}
{\sc Francois-Lovens, P.}
\newblock “without these conversations with the eliza chatbot, my husband
  would still be here”.
\newblock {\em La Libre\/}.

\bibitem{fundenberg1990evolution}
{\sc Fundenberg, D., and Maskin, E.}
\newblock Evolution and cooperation in noisy repeated games.
\newblock {\em The American Economic Review 80}, 2 (1990), 274--279.

\bibitem{gabriel2020artificial}
{\sc Gabriel, I.}
\newblock Artificial intelligence, values, and alignment.
\newblock {\em Minds and machines 30}, 3 (2020), 411--437.

\bibitem{gurouglu2009fairness}
{\sc G{\"u}ro{\u{g}}lu, B., van~den Bos, W., and Crone, E.~A.}
\newblock Fairness considerations: increasing understanding of intentionality
  during adolescence.
\newblock {\em Journal of experimental child psychology 104}, 4 (2009),
  398--409.

\bibitem{hadfield2019incomplete}
{\sc Hadfield-Menell, D., and Hadfield, G.~K.}
\newblock Incomplete contracting and ai alignment.
\newblock In {\em Proceedings of the 2019 AAAI/ACM Conference on AI, Ethics,
  and Society\/} (2019), pp.~417--422.

\bibitem{hadfield2016cooperative}
{\sc Hadfield-Menell, D., Russell, S.~J., Abbeel, P., and Dragan, A.}
\newblock Cooperative inverse reinforcement learning.
\newblock {\em Advances in neural information processing systems 29\/} (2016).

\bibitem{harrington2005theory}
{\sc Harrington, L., Siegert, R., and McClure, J.}
\newblock Theory of mind in schizophrenia: a critical review.
\newblock {\em Cognitive neuropsychiatry 10}, 4 (2005), 249--286.

\bibitem{hooker2008mentalizing}
{\sc Hooker, C.~I., Verosky, S.~C., Germine, L.~T., Knight, R.~T., and
  D’Esposito, M.}
\newblock Mentalizing about emotion and its relationship to empathy.
\newblock {\em Social cognitive and affective neuroscience 3}, 3 (2008),
  204--217.

\bibitem{kinderman1998theory}
{\sc Kinderman, P., Dunbar, R., and Bentall, R.~P.}
\newblock Theory-of-mind deficits and causal attributions.
\newblock {\em British journal of Psychology 89}, 2 (1998), 191--204.

\bibitem{knobe2005theory}
{\sc Knobe, J.}
\newblock Theory of mind and moral cognition: Exploring the connections.
\newblock {\em Trends in cognitive sciences 9}, 8 (2005), 357--359.

\bibitem{kosinski2023theory}
{\sc Kosinski, M.}
\newblock Theory of mind may have spontaneously emerged in large language
  models.
\newblock {\em arXiv preprint arXiv:2302.02083\/} (2023).

\bibitem{krakovna2020gaming}
{\sc Krakovna, V., Uesato, J., Mikulik, V., Rahtz, M., Everitt, T., Kumar, R.,
  Kenton, Z., Leike, J., and Legg, S.}
\newblock Specification gaming: the flip side of ai ingenuity, 21 April 2020.

\bibitem{kreps1982rational}
{\sc Kreps, D.~M., Milgrom, P., Roberts, J., and Wilson, R.}
\newblock Rational cooperation in the finitely repeated prisoners' dilemma.
\newblock {\em Journal of Economic theory 27}, 2 (1982), 245--252.

\bibitem{lane2010theory}
{\sc Lane, J.~D., Wellman, H.~M., Olson, S.~L., LaBounty, J., and Kerr, D.~C.}
\newblock Theory of mind and emotion understanding predict moral development in
  early childhood.
\newblock {\em British Journal of Developmental Psychology 28}, 4 (2010),
  871--889.

\bibitem{lee2021learning}
{\sc Lee, C., and Kurumada, C.}
\newblock Learning maximum absolute meaning through reasoning about speaker
  intentions.
\newblock {\em Language Learning 71}, 2 (2021), 326--368.

\bibitem{littler2015cognitive}
{\sc Littler, K.}
\newblock Cognitive and affective processes associated with moral reasoning,
  and their relationship with behaviour in typical development.

\bibitem{ma2015lie}
{\sc Ma, F., Evans, A.~D., Liu, Y., Luo, X., and Xu, F.}
\newblock To lie or not to lie? the influence of parenting and theory-of-mind
  understanding on three-year-old children’s honesty.
\newblock {\em Journal of Moral Education 44}, 2 (2015), 198--212.

\bibitem{malle2004mind}
{\sc Malle, B.~F.}
\newblock How the mind explains behavior.
\newblock {\em Folk explanation, Meaning and social interaction. Massachusetts:
  MIT-Press\/} (2004).

\bibitem{mcdonald2006reliability}
{\sc McDonald, S., Bornhofen, C., Shum, D., Long, E., Saunders, C., and
  Neulinger, K.}
\newblock Reliability and validity of the awareness of social inference test
  (tasit): a clinical test of social perception.
\newblock {\em Disability and rehabilitation 28}, 24 (2006), 1529--1542.

\bibitem{melis2010human}
{\sc Melis, A.~P., and Semmann, D.}
\newblock How is human cooperation different?
\newblock {\em Philosophical Transactions of the Royal Society B: Biological
  Sciences 365}, 1553 (2010), 2663--2674.

\bibitem{milli2017should}
{\sc Milli, S., Hadfield-Menell, D., Dragan, A., and Russell, S.}
\newblock Should robots be obedient?
\newblock {\em arXiv preprint arXiv:1705.09990\/} (2017).

\bibitem{park2023generative}
{\sc Park, J.~S., O'Brien, J., Cai, C.~J., Morris, M.~R., Liang, P., and
  Bernstein, M.~S.}
\newblock Generative agents: Interactive simulacra of human behavior.
\newblock In {\em Proceedings of the 36th Annual ACM Symposium on User
  Interface Software and Technology\/} (2023), pp.~1--22.

\bibitem{park2023ai}
{\sc Park, P.~S., Goldstein, S., O'Gara, A., Chen, M., and Hendrycks, D.}
\newblock Ai deception: A survey of examples, risks, and potential solutions.
\newblock {\em arXiv preprint arXiv:2308.14752\/} (2023).

\bibitem{perez2022discovering}
{\sc Perez, E., Ringer, S., Luko{\v{s}}i{\=u}t{\.e}, K., Nguyen, K., Chen, E.,
  Heiner, S., Pettit, C., Olsson, C., Kundu, S., Kadavath, S., et~al.}
\newblock Discovering language model behaviors with model-written evaluations.
\newblock {\em arXiv preprint arXiv:2212.09251\/} (2022).

\bibitem{powell2010orbital}
{\sc Powell, J.~L., Lewis, P.~A., Dunbar, R.~I., Garc{\'\i}a-Fi{\~n}ana, M.,
  and Roberts, N.}
\newblock Orbital prefrontal cortex volume correlates with social cognitive
  competence.
\newblock {\em Neuropsychologia 48}, 12 (2010), 3554--3562.

\bibitem{premack1978does}
{\sc Premack, D., and Woodruff, G.}
\newblock Does the chimpanzee have a theory of mind?
\newblock {\em Behavioral and brain sciences 1}, 4 (1978), 515--526.

\bibitem{ridinger2017theory}
{\sc Ridinger, G., and McBride, M.}
\newblock Theory of mind ability and cooperation.
\newblock {\em Manuscript, Univ. California, Irvine\/} (2017).

\bibitem{sally2006development}
{\sc Sally, D., and Hill, E.}
\newblock The development of interpersonal strategy: Autism, theory-of-mind,
  cooperation and fairness.
\newblock {\em Journal of economic psychology 27}, 1 (2006), 73--97.

\bibitem{schwitzgebel2023ai}
{\sc Schwitzgebel, E.}
\newblock Ai systems must not confuse users about their sentience or moral
  status.
\newblock {\em Patterns 4}, 8 (2023).

\bibitem{shamay2009two}
{\sc Shamay-Tsoory, S.~G., Aharon-Peretz, J., and Perry, D.}
\newblock Two systems for empathy: a double dissociation between emotional and
  cognitive empathy in inferior frontal gyrus versus ventromedial prefrontal
  lesions.
\newblock {\em Brain 132}, 3 (2009), 617--627.

\bibitem{shapira2023clever}
{\sc Shapira, N., Levy, M., Alavi, S.~H., Zhou, X., Choi, Y., Goldberg, Y.,
  Sap, M., and Shwartz, V.}
\newblock Clever hans or neural theory of mind? stress testing social reasoning
  in large language models.
\newblock {\em arXiv preprint arXiv:2305.14763\/} (2023).

\bibitem{shevlin2021uncanny}
{\sc Shevlin, H.}
\newblock Uncanny believers: Chatbots, beliefs, and folk psychology.
\newblock {\em Unpublished manuscript\/} (2021).

\bibitem{stevens2004nice}
{\sc Stevens, J.~R., and Hauser, M.~D.}
\newblock Why be nice? psychological constraints on the evolution of
  cooperation.
\newblock {\em Trends in cognitive sciences 8}, 2 (2004), 60--65.

\bibitem{stiller2007perspective}
{\sc Stiller, J., and Dunbar, R.~I.}
\newblock Perspective-taking and memory capacity predict social network size.
\newblock {\em Social Networks 29}, 1 (2007), 93--104.

\bibitem{sullivan1995children}
{\sc Sullivan, K., Winner, E., and Hopfield, N.}
\newblock How children tell a lie from a joke: The role of second-order mental
  state attributions.
\newblock {\em British journal of developmental psychology 13}, 2 (1995),
  191--204.

\bibitem{sutton1999bullying}
{\sc Sutton, J., Smith, P.~K., and Swettenham, J.}
\newblock Bullying and ‘theory of mind’: A critique of the ‘social skills
  deficit’view of anti-social behaviour.
\newblock {\em Social development 8}, 1 (1999), 117--127.

\bibitem{sutton1999social}
{\sc Sutton, J., Smith, P.~K., and Swettenham, J.}
\newblock Social cognition and bullying: Social inadequacy or skilled
  manipulation?
\newblock {\em British journal of developmental psychology 17}, 3 (1999),
  435--450.

\bibitem{takagishi2010theory}
{\sc Takagishi, H., Kameshima, S., Schug, J., Koizumi, M., and Yamagishi, T.}
\newblock Theory of mind enhances preference for fairness.
\newblock {\em Journal of experimental child psychology 105}, 1-2 (2010),
  130--137.

\bibitem{ullman2023large}
{\sc Ullman, T.}
\newblock Large language models fail on trivial alterations to theory-of-mind
  tasks.
\newblock {\em arXiv preprint arXiv:2302.08399\/} (2023).

\bibitem{van2021deficiencies}
{\sc van Neerven, T., Bos, D.~J., and van Haren, N.~E.}
\newblock Deficiencies in theory of mind in patients with schizophrenia,
  bipolar disorder, and major depressive disorder: A systematic review of
  secondary literature.
\newblock {\em Neuroscience \& Biobehavioral Reviews 120\/} (2021), 249--261.

\bibitem{wang2023survey}
{\sc Wang, L., Ma, C., Feng, X., Zhang, Z., Yang, H., Zhang, J., Chen, Z.,
  Tang, J., Chen, X., Lin, Y., Zhao, W.~X., Wei, Z., and Wen, J.-R.}
\newblock A survey on large language model based autonomous agents, 2023.

\bibitem{warneken2006altruistic}
{\sc Warneken, F., and Tomasello, M.}
\newblock Altruistic helping in human infants and young chimpanzees.
\newblock {\em science 311}, 5765 (2006), 1301--1303.

\bibitem{weidinger2021ethical}
{\sc Weidinger, L., Mellor, J., Rauh, M., Griffin, C., Uesato, J., Huang,
  P.-S., Cheng, M., Glaese, M., Balle, B., Kasirzadeh, A., et~al.}
\newblock Ethical and social risks of harm from language models.
\newblock {\em arXiv preprint arXiv:2112.04359\/} (2021).

\bibitem{wellman1988young}
{\sc Wellman, H.~M., and Bartsch, K.}
\newblock Young children's reasoning about beliefs.
\newblock {\em Cognition 30}, 3 (1988), 239--277.

\bibitem{wellman2008including}
{\sc Wellman, H.~M., and Miller, J.~G.}
\newblock Including deontic reasoning as fundamental to theory of mind.
\newblock {\em Human Development 51}, 2 (2008), 105--135.

\bibitem{zhao2023survey}
{\sc Zhao, W.~X., Zhou, K., Li, J., Tang, T., Wang, X., Hou, Y., Min, Y.,
  Zhang, B., Zhang, J., Dong, Z., et~al.}
\newblock A survey of large language models.
\newblock {\em arXiv preprint arXiv:2303.18223\/} (2023).

\end{thebibliography}

\end{document}